\documentstyle[11pt,moriond,epsfig]{article}

\def\Journal#1#2#3#4{{#1} {\bf #2}, #3 (#4)}

\def\PLB{{\em Phys. Lett.}  B}
\def\PRL{\em Phys. Rev. Lett.}
\def\PRD{{\em Phys. Rev.} D}
\def\PR{\em Phys. Rev.}

\def\ra{\rightarrow}

\def\mhp{$m_{H^{\pm}}$}
\def\hp{$H^\pm$}
\def\htn{$H^\pm\rightarrow\tau\nu$}
\def\ttbar {$t\bar{t}$}

\def \gev   {\rm GeV}

\def \gevcc {\rm GeV/c$^2$}

\def \ee   {${\mathrm e}^+{\mathrm e}^-$}

\def \CDF  {{\sc CDF}}
\def \dzero {{\sc{D}${\mathsc {\emptyset}}$}}
\def \LEP    {{\sc LEP}}

\def \CMS    {{\sc CMS}}
\def \ATLAS     {{\sc ATLAS}}
\def \LHC  {{\sc LHC}}
\def \QCD {{\sc QCD}}

\def\be{\begin{equation}}
\def\ee{\end{equation}}
\def\bea{\begin{eqnarray}}
\def\eea{\end{eqnarray}}

\begin{document}
\vspace*{4cm}
\title{DISCOVERY POTENTIAL FOR ${\boldmath H^\pm}$ AT LHC IN ${\boldmath H^\pm\ra\tau^\pm\nu}$  DECAYS}

\author{A. TRICOMI}

\address{Dipartimento di Fisica e Astronomia and INFN Catania\\
Catania, Italy}

\maketitle\abstracts{
The discovery potential of charged Higgs from $pp\rightarrow tH^\pm$ in the 
$H^\pm\rightarrow\tau\nu$ decay channel 
is investigated in \CMS\ and \ATLAS. For $m_{H^{\pm}}>m_t$, the 
most relevant channels are $H^\pm\ra tb$ and $H\pm\ra\tau\nu$. Whereas the former has the largest 
branching ratio it suffers of large irreducible backgrounds, while the latter offers a very clean 
enviroment when appropriate cuts are used. Making use of the $\tau$ polarization effects, in the 
purely hadronic final states an almost background-free signal is selected. The expected discovery 
range is evaluated for \CMS\ and \ATLAS\ with 30~fb$-1$ each in the low luminosity running conditions 
and the combined results are presented.}   

\section{Introduction}
\label{intro}

In the Minimal Supersymmetric Standard Model (MSSM), the Higgs sector consists of five physical 
states, two of which are charged ($H^\pm$), while the other are neutral 
($h^0$, $H^0$ and $A^0$). At tree level all Higgs masses and couplings are expressed in terms of 
two parameters, generally taken as $m_A$, the mass of the CP-odd neutral Higgs $A^0$,  
and $\tan\beta$, the ratio of the vacuum expectation values of the Higgs doublets~\cite{higgs}. 
While any of the neutral Higgs bosons may be difficult to distinguish from the Standard Model one, the 
$H^\pm$ carries a distinctive signature of the SUSY Higgs sector. Moreover the coupling of $H^\pm$ are 
uniquely related to $\tan\beta$, therefore the detection of $H^\pm$ and the measurement of its mass and couplings 
should play a very important role in probing the MSSM Higgs sector.

In a model independent way, \LEP\ experiments have set lower limits on the mass of the charged Higgs boson, 
$m_{H^{\pm}}>78.5$~\gevcc\ for any $\tan\beta$~\cite{lep}. \CDF\ and \dzero\   have performed several searches 
for \hp~\cite{tevatron,cdf}: a direct search through the decay \htn\ was performed at high $\tan\beta$ ($>10$), 
while at low $\tan\beta$ ($<1$) an indirect search for the decay channel $H^\pm\rightarrow c\bar{s}$ was 
carried out.  \dzero\ performed a dissapearance search considering all the fermionic decay modes of the 
charged Higgs~\cite{D0}. The searches at {\sc TEVATRON} excluded the low ($<1$) and 
high ($>40$) $\tan\beta$ region up to 120~\gevcc\ and 160~\gevcc, respectively.

In the study described herein, the charged Higgs discovery potential in \htn\ decay mode for \mhp $>m_t$ at \LHC\ 
is  investigated. The $t \rightarrow bH^+$ decay is known to provide promising
signatures for charged Higgs boson search at {\sc TEVATRON} upgrade and LHC
for \mhp $< m_t$~\cite{lessmt}.  But it is hard to extend the $H^\pm$ search beyond $m_t$, because in this
case the combination of dominant production and decay channels, $tH^-\rightarrow t\bar tb$, suffers from a 
large QCD background~\cite{barger}. Moreover the subdominant production channels of $H^\pm W^\mp$ and
$H^\pm H^\mp$ have been found to give no viable signature at LHC~\cite{barrentius}. 
In view of this both \ATLAS\ and \CMS\ have performed  several studies of the most promising 
decays channels of a 
heavy charged Higgs signature at \LHC~\cite{heavy}. 
Above the top quark mass the dominant decay mode is the $tb$ channel. 
However, it suffers of large irreducible background, while the $\tau\nu$ decay channel, which is the sub-dominant 
decay mode, can offer a free of background enviroment, taking advantage of the distinctive $\tau$ polarization. 
In the following, some details of the current analysis performed by \CMS\ and \ATLAS\ are discussed in the 
framework of the discovery potential achievable at \LHC.
  
\section{Analyses}

Both \CMS\ and \ATLAS\ analyses are looking for $H^\pm\rightarrow\tau\nu\rightarrow h + E_t^{miss} + X$ decays 
in which the $H^\pm$ is produced in association with a 
top quark through the processes $g b\rightarrow t H^\pm$ and $g g\rightarrow t b H^\pm$. 
The production 
cross section at \LHC\ is evaluated in ref.~\cite{heavy} eliminating the double counting for the two processes.  
The associated top quark is required to decay hadronically, $t\rightarrow jjb$, in order that the missing energy gets 
contribution mainly due to the neutrino from \htn\ and the transverse mass distribution reconstructed from the $\tau$ 
jet and the missing transverse energy, has a jacobian type structure with an endpoint at \mhp. It is also shown that 
effect of the $\tau$ polarization can be used to enhance the signal over background 
with appropriate $\tau$ selection 
cuts~\cite{heavy}. The main background, in fact, are due to the \ttbar\ events with $W_1\rightarrow\tau\nu$, 
$W_2\rightarrow qq^{\prime}$, $W+jet$ events with $W\rightarrow\tau\nu$ and $Wt$ events with  $W_1\rightarrow\tau\nu$, 
$W_2\rightarrow qq^{\prime}$. Since all these backgrounds contain the 
$W\rightarrow\tau\nu$ decay they can be efficiently 
reduced thanks to the $\tau$ polarization effect: indeed due to the scalar nature of 
the decaying $H^+$ the $\tau^+$ from 
its decay is produced in a left-handed polarization state. In the simplest scenario 
of $\tau^+\rightarrow\pi^+\bar\nu$ 
decay the right-handed $\bar\nu$ is preferentially emitted in the direction opposite 
to the $\tau^+$ in the 
$\tau$ rest frame to preserve the polarization. On the contrary, in the dominant $t\bar{t}$ 
background the $\tau^+$ from 
$W^+\rightarrow\tau^+\nu$ decay is produced right-handed due to the vector 
nature of the $W^+$ forcing the $\bar\nu$ from  
$\tau^+\rightarrow\pi^+\bar\nu$ to be emitted in the same direction as the 
$\tau^+$ in the $\tau$ rest frame. 
Therefore harder pions are expected from the signal than from the background. \\
The $\tau^\pm\rightarrow\pi^\pm\bar\nu$ contributes as 12.5\% to the charged one prong 
decays. Significant 
contributions to the one prong decay come also from vector meson production, 
$\tau^\pm\rightarrow\rho^\pm\bar\nu$ and 
$\tau^\pm\rightarrow a_1^\pm\bar\nu$ with branching ratios of 26\% and 7.5\%, respectively. In this case harder 
pions are produced from the longitudinal vector meson components relative to the background while the opposite is true 
for the transverse components, which contribute to smear the polarization effect.  This difference in the pion 
spectra can be exploited by requiring a large fraction of the $\tau$ jet energy to be carried by a single charged 
hadron in the jet. The signal selection relies heavily on the these two future for both the two analyses. To further 
reduce the large background coming from $W+jet$ events with $W\rightarrow\tau\nu$,
  $W$ and top mass recostruction is 
performed. Other cuts characteristic of each analysis is described in the following sections.

\subsection{CMS analysis}

As mentioned in the previous section, the hadronic $\tau$ signature of a heavy
charged Higgs boson from $pp \ra tH^{\pm}$ at the LHC is useful. To select signal over background 
we exploit the $\tau$ polarization effects and large missing energy. Several other cuts are used to further 
reduce the background. Details of this analysis can be found in ref.~\cite{cmsnote}. 

\begin{figure}[htb]
\begin{center}
\psfig{figure=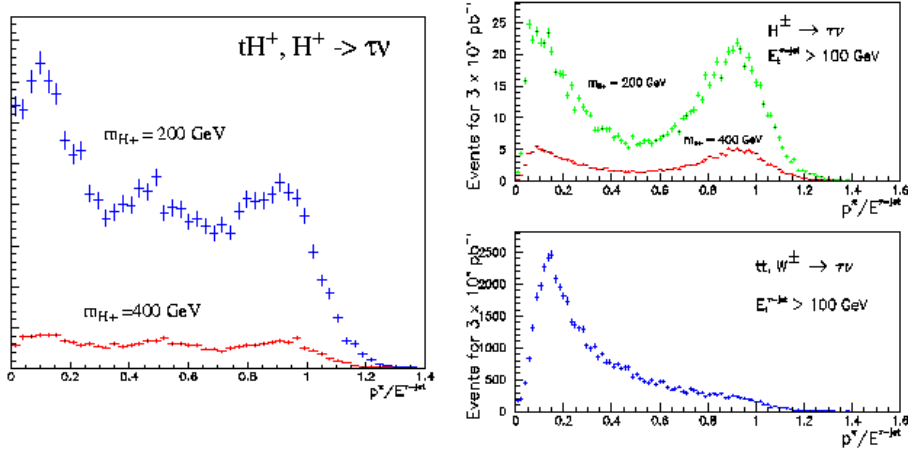,width=12truecm,clip=}
\end{center}
\caption{Distribution of $r=p^\pi/E^{\tau jet}$ for $H^\pm\ra\tau\nu$, 
\mhp=200~\gev\ and \mhp=400~\gev, (a) all hadronic $\tau$ decays; 
(b) $H^\pm,\,\tau^\pm\ra\pi^\pm\bar{\nu}$ decays are compared with  
 $W^\pm\ra\tau\nu$, $\tau^\pm\ra\pi^\pm\bar{\nu}$ from $t\bar{t}$ decays.}
\label{cms:taupol}
\end{figure}

The real $\tau$ jet is chosen as the $\tau$ jet candidate requiring $E_t>$ 100
GeV and $|\eta|<$2.5. 
To benefit of the $\tau$ polarization effect the variable 
r = $p^{\pi}$ / $E^{\tau jet} >$ 0.8 is used,
where $p^{\pi}$ is the momentum of a hard pion from $\tau$ decay in a cone of
$\Delta R <$ 0.1 around the calorimeter jet axis and $E^{\tau jet}$ is the
hadronic energy of the $\tau$ jet ($E_t>$ 100 GeV) reconstructed in the
calorimeters (electromagnetic and hadronic) in a cone of $\Delta R <$ 0.4. 
Figure~\ref{cms:taupol} shows the variable $r$ for all the hadronic $\tau$ 
decays for $m_{H^\pm}=200$ and 400~\gev\ (Fig.1a) while in  Fig.1b the  
\htn, $\tau^\pm\rightarrow\pi^\pm\bar\nu$ decays are compared with $W^\pm\ra\tau\nu$, 
$\tau^\pm\rightarrow\pi^\pm\bar\nu$ from \ttbar\ decays.  
 The efficiency of this $\tau$ selection for the signal events is about 18\% while for the
$t\overline{t}$ events the efficiency is only 0.4\% (including the $E_t$
threshold for jet). A reconstruction efficiency of 95\% is assumed for the 
hard isolated track from $\tau$.  

In Table~\ref{tab1} a complete list of all the selection perfomed is reported together with 
the efficiency for signal and background. As expected a large impact is due to the overall 
$\tau$ selection, the missing transverse energy cut and $W$ and top mass recostruction. 

A large missing transverse energy is expected in the signal events due to the
neutrino from $H^{\pm}$ decay. Efficiency
of the cut $E_t^{miss} >$ 100 GeV is about 75\% for the signal events and about
38\% for the $t\overline{t}$ background.  

For the reconstruction of the $W$ and top masses the 
events with at least three jets with $E_t >$ 20 GeV, in
addition to the $\tau$ jet, are selected. The $W$ and top masses are
reconstructed minimizing the variable $\chi^2$ = $(m_{jj}-m_{W})^2
+(m_{jjj}-m_{t})^2$, where $m_{W}$ and $m_{t}$ are the nominal $W$ and top
masses.  

\begin{table}[htb]
\caption{Efficiency for selection cuts for two different signal, (\mhp=200~\gev, $\tan\beta=15$) 
and (\mhp=400~\gev, $\tan\beta=23$) and main background decay channels.}\label{tab1}
\vspace{0.4cm}
\begin{center}
\begin{small}
\begin{tabular}{|c|c|c|c|c|c|}
\hline
 & (\mhp=200,$\tan\beta=15$) & (\mhp=400,$\tan\beta=23$) & $t\bar{t}$ & $Wtb$ & $W+jet$ \\
\hline
$E_t>100$ GeV & 28.1\% & 65.4\% & 8.2\% & 3.9\% & 3.7\% \\
$r>0.8$       & 26.4\% & 27.4\% & 5.2\% & 5.9\% & 9.6\% \\
total $\tau$ selection & 7.4\% & 17.9\% & 0.4\% & 0.2\% & 0.36\% \\
\hline
$E_t^{miss}>100$ GeV & 28.7\% & 74.7\% & 37.6\% & 28.4\% & 42.1\% \\
\hline
$W$ and top mass & & & & & \\
reconstructuion & 42.7\% & 38.9\% & 46.1\% & 30.9\%& 6.5\% \\
\hline
b-tagging & 50.0\% & 50.0\% & 50.0\% & 50.0\% & 1.3\% \\
\hline
Second top veto & 87.7\% & 95.6\% & 47.0\% & 77.5\% & 75.3\% \\
\hline
$\Delta\phi(\tau^{jet},E_t^{miss})>60^\circ$ & 53.2\% & 90.3\% & & & \\
\hline
\end{tabular}
\end{small}
\end{center}
\end{table}


To further reduce the background,  
after the $W$ and top mass reconstruction and the mass window cuts, b-tagging is
applied on the jet not assigned to the $W$. This jet is required to be harder
with $E_t^{jet}>$ 30 GeV. The tagging efficiencies based on the impact
parameter method obtained from a full simulation and track reconstruction in
the CMS tracker are used~\cite{tracker}. For b-jets with $E_t$ = 50 GeV the efficiency is
found to be $\sim$ 50\% averaged over the full $\eta$ range ($|\eta|<2.5$). The
mis-tagging rate for the corresponding light quark and gluon jets is 1.3\%.


The reconstructed transverse mass $m_T^{\tau\nu}$ over the total background is
shown in Fig.~2a for $m_{H^{\pm}}$ = 400 GeV and $\tan\beta$ = 40 for 30
fb$^{-1}$. For $m_T^{\tau\nu}>$100 GeV about 40 signal events are expected. 
About 5 background events from $t\overline{t}$ and $W+jet$ are 
expected for $m_T^{\tau\nu}>$100 GeV. 

Further reduction of the $t\overline{t}$ background is still
possible using a jet veto cut and a veto on a second top in the event. The
central jet veto and the second top veto, being closely correlated cuts, reduce
$t\overline{t}$ background by a factor of $\sim$7. 
The transverse mass $m_T^{\tau\nu}$ distribution over the total
background including the jet and second top veto is shown in Fig.~2b for
$m_{H^{\pm}}$ = 400 GeV and $\tan\beta$ = 40.

\begin{figure}[htb]
\begin{center}
\psfig{figure=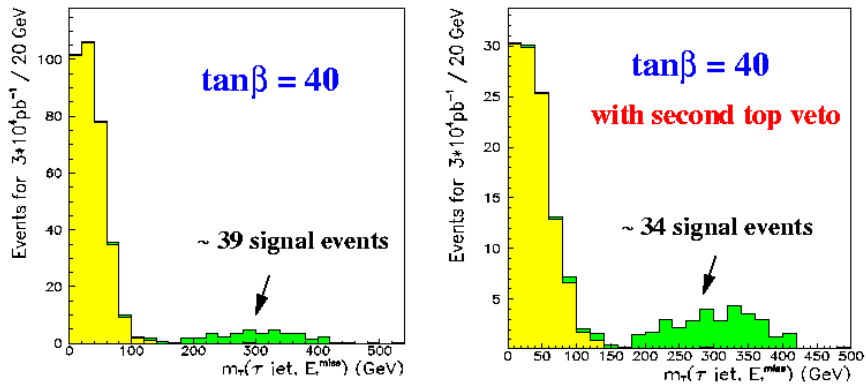,width=12truecm,clip=}
\end{center}
\caption{a) Transverse mass reconstructed from $\tau$ jet and $E_t^{miss}$ for 
$H^{\pm}\ra \tau\nu$ from $pp\ra tH^{\pm}$ with $m_{H^{\pm}}$ = 400 GeV and 
$\tan\beta$ = 40 over the total background from $t\overline{t}$ and $W$+jet 
events with basic selection cuts. b) the same as in a) but with second top veto.}
\label{mt2}
\end{figure}

The visibility of the signal can be significantly improved, especially at
$m_{H^{\pm}}$ = 200 GeV, with a cut on the $\Delta\phi$ angle between the
$\tau$ jet and the $E_t^{miss}$. Although $\Delta\phi$ is directly proportional
to $m_T^{\tau\nu}$, a cut in $\Delta\phi$ suppresses the background efficiently
at the lower end of the expected signal region.

After all the cuts a significant signal is expected for $\tan\beta>15$ between 
180~\gev$\le m_A\le 200$~\gev\ for an integrated luminosity of 30~pb$^{-1}$. 
For $m_A<200$~\gev\ the branching ratio for \htn\ 
increases rapidly however the selection efficiency, especially the one of the 
$\Delta\phi$ cut, deacreases significantly. The sensitivity for a heavy Higgs 
is for $\tan\beta>25$ at $m_A\sim 400$~\gev\ and for $\tan\beta>42$ at 
$m_A=600$~\gev. Study of the observability of \htn\ decay in the high luminosity 
running conditions are in progress, as well as the effect of stop mixing and lighter 
SUSY scale on the cross sections and branching ratios.

\subsection{ATLAS analysis}

\begin{figure}[htb]
\begin{center}
\psfig{figure=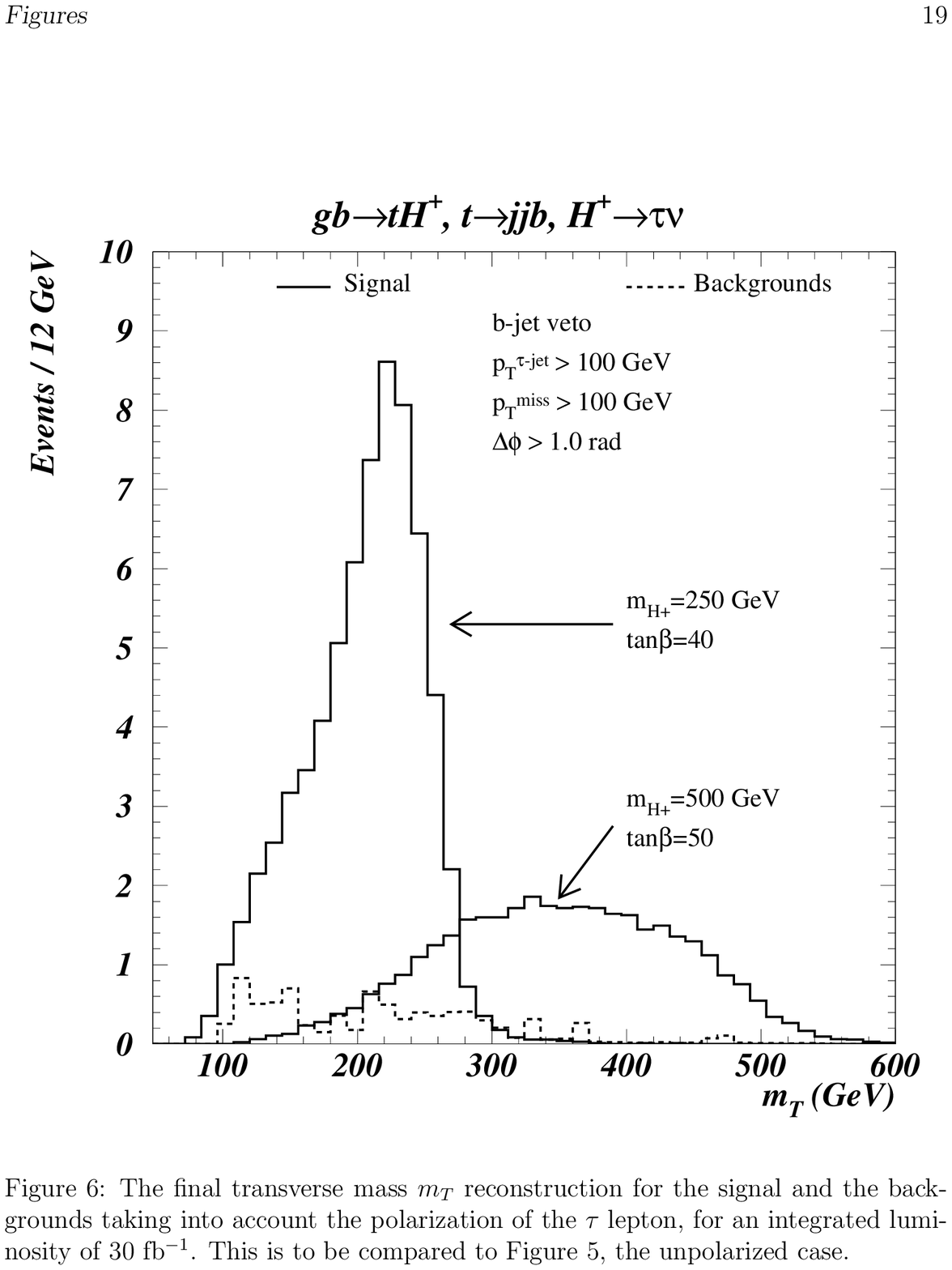,width=65truemm,clip=}
\end{center}
\caption{The final transverse mass $m_T$ reconstruction for signal and background, for 
an integrated luminosity of 30~pb$^{-1}$.}
\label{fig:atlas}
\end{figure}

The strategy used by \ATLAS\ is similar to the one of \CMS. Details can be found 
in ref.~\cite{atlasnote}. 
Several different cuts are used to 
separate signal from background exploiting the $\tau$ polarization effect and the large missing 
energy due to the neutrino from \htn.  Advantage is taken of the different kinematical 
properties of the signal and background: imposing a large cut on the $p_T$ of $\tau$ jet, the 
background events need 
a large boost from $W$ boson to satisfy this cut. This results in a small azimuthal 
opening angle  between the $\tau$ jet and the missing momentum while, on the 
contrary, the signal events not require such a boost, leading to a backward peak in the azimuthal opening 
angle. The difference in azimuthal angle between signal and background and in the missing momentum, which 
increases with \mhp, can be taken into account looking at the transverse mass 
distribution. For the signal 
the transverse mass is bound from above to \mhp, while for the background the 
transverse mass is constrained 
to be lower than $m_W$.  $W$ and top quark mass reconstruction is performed to 
further reduce $W+jet$ background  
and b tagging and second b-jet veto are also used to improve the signal selection. 
The final transverse mass 
reconstruction for signal and background is shown in Fig.~\ref{fig:atlas}. 
An almost background free signal is 
selected; significances upwards of 5$\sigma$ can be achieved for \mhp$>m_t$ 
and $\tan\beta>10$, for an integrated luminosity of 30 fb$^{-1}$. 

\section{Conclusions}

\begin{figure}[htb]
\begin{center}
\mbox{\epsfig{file=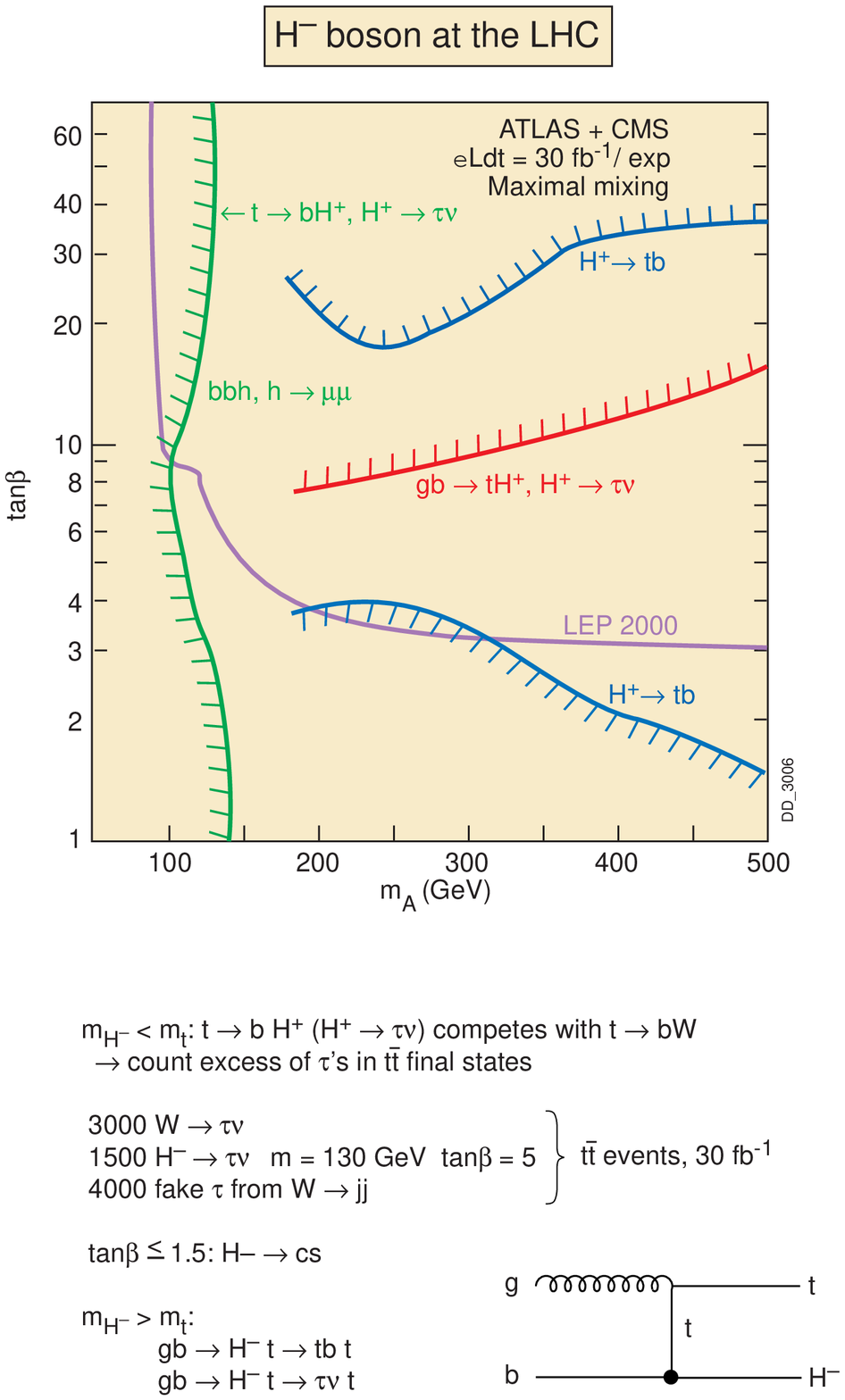,width=7truecm,bbllx=90,bblly=276,bburx=496,bbury=680,clip=}}
\end{center}
\caption{Expected 5$\sigma$ discovery limits for $pp\ra tH^\pm,
\,H^\pm\ra \tau\nu$ and $t\ra jjb$, for an integrated luminosity of 
30~fb$^{-1}$ for each experiment. Limits for other $H^\pm$ decay 
channels and the parameter space excluded by \LEP\ 
experiments in the no-mixing scenario are also shown in the figure.}  
\label{fig:5sigma}
\end{figure}

The preliminary studies made by \ATLAS\ and \CMS\ leads to the conclusion that 
$H^{\pm}\ra \tau\nu$ from $pp\ra tH^{\pm}$ is a very promising discovery channel 
for charged Higgs bosons at the \LHC. For \mhp$>m_t$, the $tb$ and the $\tau\nu$ 
decay channels constitute the main decay modes of the charged Higgs. While the 
first suffers of a large irreducible \QCD\ background, the latter could be free 
of such background, thereby extending the discovery potential beyond that 
achievable in the $tb$ channel, especially at large $\tan\beta$. 
It has been demonstrated that in the purely hadronic final states 
an almost background free signal is selected benefitting of the 
$\tau$ polarization effect and the large  missing energy due to the 
neutrino from \htn\ decay. As it is shown in Fig.~\ref{fig:5sigma}, 
a combined significance upwards of 5$\sigma$ can be achieved 
in a large part of the parameter space ($\tan\beta>10$ for \mhp$>m_t$), 
for an integrated luminosity of 30~fb$^{-1}$ for each experiment. 
Indeed, the range of discovery potential is limited by the 
signal size itself. 

\section*{Acknowledgments}
I would like to thank R. Kinnunen and D. Denegri for very helpful discussions 
and K. Assamagan for provide me useful material.  
I also thank the conference organizers for their kind and friendly hospitality.


\section*{References}
\begin{thebibliography}{99}
\bibitem{higgs} P. H. Nilles, \Journal{\PR}{110}{1}{1984}; H. Haber and G. Kane, \Journal{\PR}{115}{75}{1985}, \\
J.F. Gunion, H.E. haber, G.L. Kane and S. Dawson in {\em The Higgs Hunters' Guide} (Addison-Wesley, reading, MA, 1990).

\bibitem{lep} A. Holzner, {\em Search for Charged Higgs Bosons at LEP}, XXXVI Rencontres de Moriond, 
Electroweak Interactions and Unified Theories, Les Arcs, France, March 2001.

\bibitem{tevatron} L. Groer for \CDF\ and \dzero\ Collaborations, hep-ex/9707034.

\bibitem{cdf} \CDF\ Collaboration, \Journal{\PRL}{79}{357}{1997}.

\bibitem{D0} \dzero\ Collaboration, hep-ex/9902028.

\bibitem{lessmt} S. Raychaudhuri and D.P. Roy, \Journal{\PRD}{52}{1556}{1995} 
and \Journal{\PRD}{53}{4902}{1996}; E. Ma {\it et al}, \Journal{\PRL}{80}{1162}{1998}.

\bibitem{barger} V. Barger {\it et al}, \Journal{\PLB}{324}{236}{1994}; J.F. Gunion, 
\Journal{\PLB}{322}{125}{1994}.

\bibitem{barrentius} A.A. Barrientos Bendez\'u and B.A. Kniehl, \Journal{\PRD}{59}{015009}{1999} 
and hep-ph/9908385; S. Moretti and K. Odagiri, \Journal{\PRD}{59}{055008}{1999}. 

\bibitem{heavy} D.P. Roy, \Journal{\PLB}{459}{607}{1999}; S. Moretti and D.P. Roy, \Journal{\PLB}{470}{209}{1999}; 
M. Dress {\it et al}, \Journal{\PLB}{471}{39}{1999}; D.P. Roy, hep-ph/0102091; K.A. Assamagan, \ATLAS\ 
Internal Note ATL-PHYS-99-013, ATL-PHYS-99-025 (1999).

\bibitem{cmsnote} R. Kinnunen, \CMS\ Internal Note CMS NOTE 2000/045 (2000).

\bibitem{tracker} CMS Collaboration, The tracker project, Technical Design 
Report, CERN/LHCC 98-6, CMS TDR 5, 26 February 1998.

\bibitem{atlasnote} K.A. Assamagan, \ATLAS\ Internal Note ATL-PHYS-2000-031 (2000).

\end{thebibliography}
\end{document}